\pdfoutput=1
\documentclass[runningheads]{llncs}
\usepackage{url}
\usepackage{subfig}
\usepackage{booktabs}
\usepackage{graphicx}
\usepackage{hyperref}

\usepackage{bbding}
\usepackage{algorithm}
\usepackage[fleqn]{amsmath}
\usepackage{mathtools}
\usepackage{amsfonts,amssymb}
\usepackage{pifont}% http://ctan.org/pkg/pifont

\usepackage[nocompress]{cite}
\usepackage{booktabs} 
\usepackage[normalem]{ulem}
\usepackage{graphicx}
\usepackage{amsmath}
\usepackage{amssymb}
\usepackage{color}
\usepackage{algorithmic,algorithm}
\usepackage{hyperref}
\usepackage{multirow}
\usepackage{mathtools}
\usepackage{enumitem}
\useunder{\uline}{\ul}{}
\usepackage{threeparttable}
\usepackage{todonotes}
\usepackage{adjustbox}

\definecolor{applegreen}{rgb}{0.55, 0.71, 0.0}
\newcommand{\Reals}{\mathbb{R}}
\newcommand{\classfier}{\mathit{f}}

\newcommand{\arbitraryinputa}{\mathbf{p}}
\newcommand{\arbitraryinputb}{\hat{\mathbf{p}}}

\newcommand{\image}{\mathbf{I}}
\newcommand{\segmentation}{\mathbf{y}}

\newcommand{\advnoise}{\mathbf{r}^{\rm{ adv}}}
\newcommand{\noise}{\mathbf{r}}
\newcommand{\cpoints}{\mathbf{c}}

\newcommand{\one}{\mathbf{1}}

\newcommand{\biasfunc}{\mathcal{G}_{\rm bias}(\image; \cpoints)}
\newcommand{\advbiasfunc}{\mathcal{G}_{\rm bias}(\image; \cpoints^{adv})}

\DeclareMathOperator{\proj}{\Pi} 
\newcommand{\sobel}{\textit{S}}

\newcommand{\Loss}{\mathcal{L}}
\newcommand{\SegLoss}{\mathcal{L}_\mathrm{seg}}

\newcommand{\Distance}{\mathcal{D}}
\newcommand{\CompositeDistance}{\Distance_{\rm{comp}}}

\newcommand{\KLDiv}{\Distance_\mathrm{KL}}
\newcommand{\ContourDist}{\Distance_\mathrm{contour}}
\newcommand{\weight}{w}

\title{Realistic Adversarial Data Augmentation for MR Image Segmentation}
\author{Chen Chen\inst{1}(\Envelope),  Chen Qin\inst{1}, Huaqi Qiu\inst{1}, Cheng Ouyang\inst{1}, Shuo Wang\inst{3}, Liang Chen\inst{1,4}, Giacomo Tarroni\inst{1,2},  Wenjia Bai\inst{3,4}, Daniel Rueckert\inst{1}}
\authorrunning{C. Chen et al.}

\institute{BioMedIA Group, Department of Computing, Imperial College London, UK\\
\and CitAI Research Centre, Department of Computer Science, City, University of London, UK
\and Data Science Institute, Imperial College London, UK
\and Department of Brain Sciences, Imperial College London, UK
\\
\email{chen.chen15@imperial.ac.uk}}

\titlerunning{Adversarial data augmentation for MR segmentation}
\begin{document}
\maketitle  
% https://www.miccai2020.org/en/PAPER-SUBMISSION-GUIDELINE.html#manuscript-format

% The papers will be evaluated by three external reviewers and Area Chairs for potential inclusion in the scientific program of MICCAI.
%Each paper must be submitted with Primary and Secondary areas selected from the CMT system and should also indicate up to three relevant keywords in the "keyword” section of the paper template. Authors also must identify to which stream the paper belongs, i.e., either MIC, CAI, or MICCAI. These areas, the stream, and the paper itself, will be used to generate suggested reviewers using the automated TPMS paper matching system embedded in the CMT system 
\begin{abstract}
Neural network-based approaches can achieve high accuracy in various medical image segmentation tasks. However, they generally require large labelled datasets for supervised learning. Acquiring and manually labelling a large medical dataset is expensive and sometimes impractical due to data sharing and privacy issues. In this work, we propose an adversarial data augmentation method for training neural networks for medical image segmentation. Instead of generating pixel-wise adversarial attacks, our model generates plausible and realistic signal corruptions,  which models the intensity inhomogeneities caused by a common type of artefacts in MR imaging: bias field. The proposed method does not rely on generative networks, and can be used as a plug-in module for general segmentation networks in both supervised and semi-supervised learning. Using cardiac MR imaging we show that such an approach can improve the generalization ability and robustness of models as well as provide significant improvements in low-data scenarios.  
\keywords{Image segmentation, Adversarial data augmentation, MR}
\end{abstract}

\section{Introduction}
\label{SEC:introduction}
%%State the importance of this task: 
Segmentation of medical images is an important task for diagnosis, treatment planning and clinical research~\cite{Smistad_2015_MedIA}.
%% The core challenges are ...:
Recent years have witnessed the fast development of deep learning for medical imaging with neural networks being applied to a variety of medical image segmentation tasks~\cite{Shen_2017_Review,litjens_2017_survey}. Deep learning-based approaches in general require a large-scale labelled dataset for training, in order to achieve good model generalization ability and robustness on unseen test cases. However, acquiring and manually labelling such large medical datasets is extremely challenging, due to the difficulties that lie in data collection and sharing, as well as to the high labelling costs~\cite{Tajbakhsh_2019_Arxiv}.
% several reasons. First, it is well-known that collecting and sharing data from clinical sites is difficult mainly due to data privacy and ethical issues. Second, labelling medical image usually requires expertise and thus is time-consuming and expensive. As a result, it is often the case that only a small number of labelled images are available for training, which hinders the deployment of deep learning methods in real world. 

%% Previous work have addressed .....
To address the aforementioned problems, one of the commonly adopted strategies is data augmentation, which aims to increase the diversity of the available training data without collecting and manually labelling new data. Conventional data augmentation methods mainly focus on applying simple \emph{random} transformations to labelled images. These random transformations include intensity transformations (e.g. pixel-wise noise, image brightness and contrast adjustment) and geometric transformations (e.g. affine, elastic transformations). Recently, there is a growing interest in developing generative network-based methods for data augmentation~\cite{Zhao_2019_CVPR_oneshotDA,Liu_2019_MICCAI,Chaitanya_2019_IPMI,Xing_2019_MICCAI}, which have been found effective for one-shot brain segmentation~\cite{Zhao_2019_CVPR_oneshotDA} and low-shot cardiac segmentation~\cite{Chaitanya_2019_IPMI}. Unlike conventional data augmentation, which generates new examples in an uninformative fashion and does not account for complex variations in data, this generative network-based method is data-driven, learning optimal image transformations from the underlying data distribution in the real world~\cite{Chaitanya_2019_IPMI}. However, in practice, training generative networks is not trivial due to their sensitivity to hyper-parameters tuning~\cite{lei_2020_geometric} and it can suffer from the mode collapse problem.

%% Our works.
In this work, we introduce an effective adversarial data augmentation method for medical imaging without resorting to generative networks. Specifically, we introduce a realistic intensity transformation function to amplify intensity non-uniformity in images, simulating potential image artefacts that may occur in clinical MR imaging (i.e. bias field). Our work is motivated by the observations that MR images often suffer from low-frequency intensity corruptions caused by inhomogeneities in the magnetic field. This artefact cannot be easily eliminated~\cite{Tustison_2010_N4ITK,khalili2019automatic} and can be regarded as a physical attack to neural networks, which have been reported to be sensitive to intensity perturbations~\cite{paschali_2018_MICCAI_generalizability,Chen_2019_SASHIMI_Intelligent}. To efficiently improve the model generalizability and robustness, we apply adversarial training to directly search for optimal intensity transformations that benefit model training. By continuously generating these realistic, `hard' examples, we prevent the network from over-fitting and, more importantly, encourage the network to defend itself from intensity perturbations by learning robust semantic features for the segmentation task.

%% Our contributions.
Our main contributions can be summarised as follows:
(1) We introduce a realistic adversarial intensity transformation model for data augmentation in MRI, which simulates intensity inhomogeneities which are common artefacts in MR imaging. The proposed data augmentation is complementary to conventional data augmentation methods.
(2) We present a simple yet effective framework based on adversarial training to learn adversarial transformations and to regularize the network for segmentation robustness, which can be used as a plug-in module in general segmentation networks. More importantly, unlike conventional adversarial example construction~\cite{Goodfellow_2015_FGSM, Madry_2017_PGDattack,papernot_2018_cleverhans}, generating adversarial bias fields does not require manual labels, which makes it applicable for both supervised and semi-supervised learning, see Sec.~\ref{SEC:ADV}. 
% (3) We propose a new composite regularization term that takes both label distribution information and shape information into account, which encourages the network to produce consistent and precise predictions against small perturbations (Eq.~\ref{eq: divergence loss function}).
(3) We demonstrate the efficacy of the proposed method on a public cardiac MR segmentation dataset in challenging low-data settings. In this scenario, the proposed method greatly outperforms competitive baseline methods, see Sec.~\ref{SEC: results}.

\noindent\textbf{Related work.}
Recent studies have shown that adversarial data augmentation, which generates adversarial data samples during training, is effective to improve model generalization and robustness\cite{Madry_2017_PGDattack,Volpi_2018_NIPS_Generalizing}. Most existing works are based on designing attacks with pixel-wise noise, i.e. by adding gradient-based adversarial noise~\cite{Goodfellow_2015_FGSM,Carlini_2017_CW_attack,Tramer_2019_NIPS_Adversarial,Miyato_2018_PAMI_VAT, Paschali_2018_MICCAI}. More recently, there have been studies showing that neural networks can also be fragile to other, more natural form of transformations that can occur in images, such as affine transformations~\cite{Kanbak_2018_CVPR_Geometric,Engstrom_2019_ICML,Zeng_2019_CVPR_Beyond}, illumination  changes~\cite{Zeng_2019_CVPR_Beyond}, and small deformations~\cite{Alaifari_2019_ICLR_ADef,Chen_2019_SASHIMI_Intelligent}. In medical imaging, designing and constructing realistic adversarial perturbations, which can be used for improving medical image segmentation networks, has not been explored in depth.

\section{Adversarial Data Augmentation with Robust Optimization}
\label{SEC:methodology}
In this work, we aim at generating realistic adversarial examples to improve model generalization ability and robustness, given a limited number of training examples. To achieve the goal, we first introduce a physics-based intensity transformation model that can simulate intensity inhomogeneities in MR images. We then propose an adversarial training method, which finds effective adversarial transformation parameters to augment training data, and then regularizes the network with a distance loss function which penalizes network's sensitivity to such adversarial perturbations. Since our method is based on virtual adversarial training (VAT)~\cite{Miyato_2018_PAMI_VAT}, we will first briefly review VAT before introducing our method.

\subsection{Virtual Adversarial Training}
\label{SEC:VAT}
\begin{figure}[t]
    \centering
    \includegraphics[width=\textwidth]{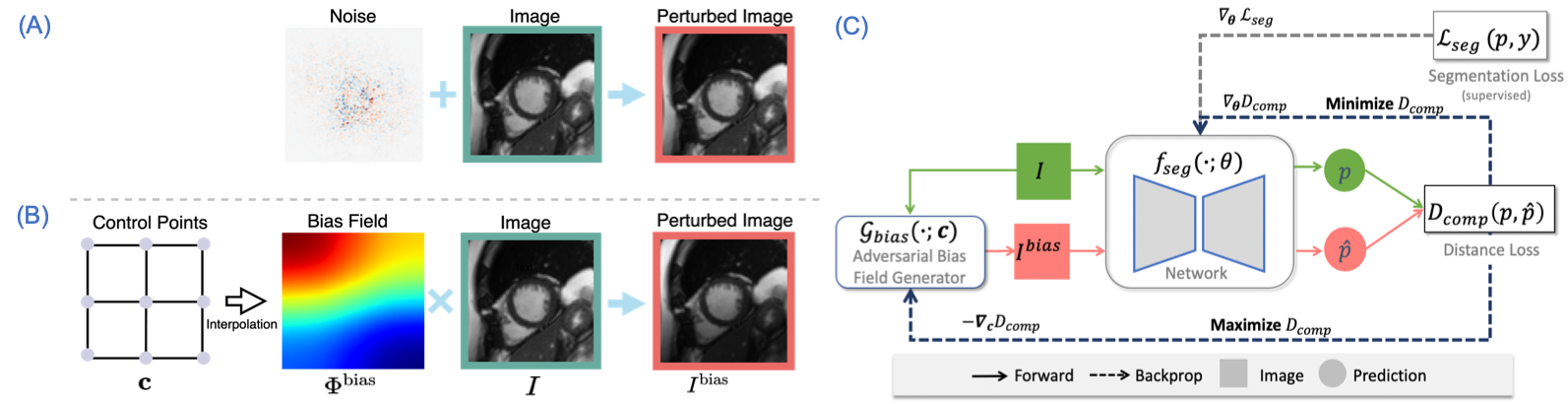}
    \caption{(A) Adversarial example construction with additive gradient-based noise in VAT~\cite{Miyato_2018_PAMI_VAT};
    (B) Adversarial example construction with a multiplicative control point-based bias field  (proposed); (C) Adversarial training with bias field perturbation.}
    \label{fig:introduction}
\end{figure}

VAT is a regularization method based on adversarial data augmentation, which can prevent the model from over-fitting and improve the generalization performance and robustness\cite{Miyato_2018_PAMI_VAT}. Given an input image $\image \in \Reals^{H \times W \times C}$ (H,W,C denote image height, width, and number of channels, respectively) and a  classification network $\classfier_{cls}(\cdot;\theta)$, VAT first finds a small adversarial noise $\advnoise \in  \Reals^{H \times W \times C} $ to construct its adversarial example $\image^{adv}=\image+\advnoise$ (as shown in Fig.\ref{fig:introduction}A), with the goal of maximising the Kullback$-$Leibler (KL) divergence~$\KLDiv$ between an original probabilistic prediction $\classfier_{cls}(\image;\theta)$ and its perturbed prediction~$\classfier_{cls}(\image+\advnoise;\theta)$. The adversarial example is then used to regularize the network for robust feature learning. 
% This algorithm can be formulated as follows: 
% \begin{equation}
% \begin{aligned}
% & \underset{\theta}{\text{minimize}}
% & & \KLDiv[\classfier_{cls}(\image;\theta)\;\|\;\classfier_{cls}(\image+\advnoise;\theta)] \\
% & {\text{where } \underset{\noise}\argmax}
% & & \KLDiv[\classfier_{cls}(\image;\theta)\;\|\;\classfier_{cls}(\image+\noise;\theta)]  \\
% & \text{subject to}
% & & {\|\noise\|\leq\epsilon}, \epsilon>0.
% \end{aligned}
% \end{equation}
 
The adversarial noise can be generated by taking the gradient of $\KLDiv$ with respect to a random noise vector: $\advnoise=\epsilon\cdot\frac{r'}{\|r'\|_2}\;$,$ r'=\nabla_{\noise}{\KLDiv[\classfier(\image;\theta)\;\|\;\classfier(\image+\noise;\theta)]}$. Here $\epsilon$ is a hyper-parameter that controls the strength of perturbation. After finding adversarial examples, one can utilize them for robust learning, which penalizes the network's sensitivity to local perturbations. This is achieved by adding $\KLDiv$ to its main objective function. 

\subsection{Adversarial Training by Modelling Intensity Inhomogeneities}
\label{SEC:ADV}
In this work, we extend the VAT approach by introducing a new type of adversarial attack, namely intensity inhomogeneities (bias field) that often occur in MR imaging.
In MR imaging, a bias field is a low frequency field that smoothly varies across images, introducing intensity non-uniformity across the anatomy being imaged. The model for the intensity non-uniformity can be defined as follows~\cite{sled_1998_TMI_nonparametricN3ITK,Tustison_2010_N4ITK}: $\image^{\rm{bias}}=\biasfunc=\image \times \Phi^{\rm bias}{(\cpoints)}$.
Here, the intensity of the image $\image$ is perturbed with a multiplication with the bias field $\Phi^{\rm{bias}} \in \Reals^{H \times W}$. As the bias field is typically composed of low frequencies and thus slowly varying across the image, it can be modelled using a set of uniformly distributed $k$ by $k$ points $\cpoints=\{\cpoints_{(i)}\}_{1...k \times k}$~\cite{Tustison_2010_N4ITK}, see Fig.~\ref{fig:introduction}B. A smooth bias field at the finest resolution is obtained by interpolating scattered control points with a third-order B-spline smoothing~\cite{Gallier_2000_curves_book}.
% %and an independent addictive white Gaussian noise. %$\n\sim \GaussianD(0,\epsilon I)$%. 
% Similar to \cite{Tustison_2010_N4ITK}, we use a set of uniformly distributed control points  $\cpoints=\{\cpoints_{(i)}\}_{1...k \times k}$, which are uniformly distributed over the image grid, to govern the strength of intensity transformation at corresponding local areas. 

% And a smooth bias field at the finest resolution can be obtained by using kernel-based interpolation (e.g. Gaussian, B-spline):$\Phi_{\rm{bias}}=S_{\rm kernel}({\cpoints})$.
% In our experiments, $S_{\rm kernel}$ is a third-order B-spline convolution kernel.
While one can repeatedly sample random bias fields for data augmentation, this might be computationally inefficient as it may generate images  which are of no added value for model optimization. 
We therefore would like to construct adversarial examples (perturbed by bias field as described above) targeting the weakness of the network in an intelligent way. This allows the use of the generated adversarial examples to improve the model performance and robustness, which can be achieved via the following min-max game:
\begin{equation}
\label{opt:bias field}
\begin{aligned}
& \underset{\theta}{\text{min}}\; \underset{\cpoints}{\text{max}}
& & \CompositeDistance[\classfier_{seg}(\image;\theta),\classfier_{seg}(\biasfunc;\theta)] \\
% & \underset{\cpoints}{\text{maximize}}
% & & \CompositeDistance[\classfier_{seg}(\image;\theta),\classfier_{seg}(\biasfunc;\theta)] \\
& \text{subject to}
&&  \forall{(x,y) \in \Reals^2},\;\Phi^{\rm bias}_{(x,y)}>0;\; |{\Phi^{\rm bias}} -\one|_\infty \leq \alpha, 0<\alpha <1.
% & &\sum_{(x,y)}\Phi_{\rm bias} \sim \sum_{(x,y)}{\one};
% &&\forall{(x,y) \in \Reals^2}, |{\Phi_{\rm bias}^{(x,y)}} -1| \leq \alpha, 0<\alpha <1
\end{aligned}
\end{equation}
As shown in Fig.~\ref{fig:introduction}C, given a segmentation network $\classfier_{seg}(\cdot; \theta)$ and an input image $\image$, we first find optimal values for control points $\cpoints$ in the search space to construct an adversarial bias field, so that it \textbf{maximizes} the distance measured by~$\CompositeDistance$ between the original prediction and the prediction after perturbation: $\arbitraryinputa=\classfier_{seg}(\image;\theta),\arbitraryinputb=\classfier_{seg}(\biasfunc;\theta)$, with $\theta$ fixed. We then optimize the parameters $\theta$ in the network to \textbf{minimize} the distance between the original prediction and the prediction after the generated adversarial bias attack $\classfier_{seg}(\advbiasfunc;\theta)$. 

% Next, we will illustrate how we find the adversarial bias field first, and then introduce the proposed loss function $\CompositeDistance$ for distance measurement.

\noindent\textbf{Finding adversarial bias fields.} 
\label{SEC:bias field optimization}
To find the optimal values for the control points~$\cpoints$ for adversarial example construction, we use the gradient descent algorithm and search the values of control points in its log space for numerical stability~\cite{sled_1998_TMI_nonparametricN3ITK,Tustison_2010_N4ITK}, which allows to produce positive bias fields. Specifically, similar to the projected gradient decent   
(PGD) attack construction in~\cite{Madry_2017_PGDattack}, we first randomly initialize the values of control points and then apply a projected gradient ascent algorithm to iteratively update $\cpoints$ with $n$ steps: $\cpoints \leftarrow \proj( \cpoints+ \xi\cdot \cpoints'/\|{\cpoints'}\|_2)$ where $\cpoints'= \nabla_\cpoints{\CompositeDistance} [\classfier_{seg}(\image;\theta),\classfier_{seg}(\biasfunc;\theta)]$. $\proj$ denotes the projection function which projects $\cpoints$ onto the feasible set, and $\xi$ is the step size. For neural networks, gradients $\cpoints'$ can be efficiently computed with back-propagation. $\Phi^{\rm{bias}}$ is updated by first interpolating the coarse-grid control points (log values at the current iteration) to its finest grid using B-spline convolution, and then taking the exponential function for value recovering. Finally, the generated bias field is rescaled to meet the magnitude constraint in Eq.~\ref{opt:bias field}. 

% \notsure{The whole procedure is summarized in Algorithm~\ref{alg:bias} in the supplementary material.}
% \todo{Not sure if MICCAI allows me to put algorithm in supplementary material.}

\noindent\textbf{Composite distance function~$\CompositeDistance$.}  Here, we propose a composite distance function $\CompositeDistance$ to enhance its discrimination ability between the original prediction $\arbitraryinputa$ (short for $\classfier_{seg}(\image;\theta)$) and the prediction after perturbation $\arbitraryinputb$, for \emph{semantic segmentation} tasks. This composite loss consists of (1) the original $\KLDiv$ used in VAT, which measures the difference between distributions and (2) a contour-based loss function $\ContourDist$ \cite{chen2019unsupervised} which is specifically designed to capture mismatch between object boundaries:
$\CompositeDistance(\arbitraryinputa,\arbitraryinputb)=\KLDiv[\arbitraryinputa\;||\;\arbitraryinputb]+\weight\ContourDist(\arbitraryinputa,\arbitraryinputb)$;
$\ContourDist(\arbitraryinputa,\arbitraryinputb)=\sum_{m \in M}\sum_{\sobel_{x,y}} \left\|\sobel({\arbitraryinputa^m})-\sobel(\arbitraryinputb^{m})\right\|_{2}$. M denotes foreground channels, $\sobel_{x,y}$ denote two Sobel filters in $x$- and $y$-direction for edge extraction and $\weight$ controls the relative importance of both terms.

\noindent\textbf{Optimizing segmentation network.}  After constructing the adversarial examples, one can compute $\CompositeDistance$ and apply it to regularizing the network, encouraging the network to be less sensitive to adversarial perturbations, and thus produce consistent predictions. Since this algorithm uses probabilistic predictions (produced by the network) rather than manual labels for adversary construction, it can be applied to both labelled ($l$) and unlabelled data ($u$) for supervised and semi-supervised learning~\cite{Miyato_2018_PAMI_VAT}. The loss functions for the two scenarios are defined as: $\Loss_{\rm{SU}}= \SegLoss (\arbitraryinputa^{(l)},\segmentation_{gt}^{(l)})+\lambda_l\CompositeDistance(\arbitraryinputa^{(l)},\arbitraryinputb^{(l)}); \; \Loss_{\rm{SE}}=\Loss_{\rm{SU}}+\lambda_u\CompositeDistance(\arbitraryinputa^{(u)},\arbitraryinputb^{(u)}).$ 
$\SegLoss$ denotes a general task-related segmentation loss function for supervised learning (e.g. cross-entropy loss) and $\segmentation_{gt}^{(l)}$ denotes ground truth.

\section{Experiments}
\label{SEC: experiment}
To test the efficacy of the proposed method, we applied it to training a segmentation network for the left ventricular myocardium from MR images in low-data settings. We compared the results with several competitive baseline methods.  
\subsection{Dataset and Experiment Settings}
\noindent\textbf{ACDC dataset.} Experiments were performed on a public benchmark dataset for cardiac MR image segmentation: The Automated Cardiac Diagnosis Challenge (ACDC) dataset~\cite{Bernard_2018_ACDC}~\footnote{\url{https://www.creatis.insa-lyon.fr/Challenge/acdc/databases.html}}. This dataset was collected from 100 subjects which were evenly classified into 5 groups: 1 normal group (NOR) and 4 pathological groups with cardiac abnormalities: dilated cardiomyopathy(DCM); hypertrophic cardiomyopathy (HCM); myocardial infarction with altered left ventricular ejection
fraction (MINF); abnormal right ventricle (ARV). The left ventricular myocardium in end-diastolic and end-systolic frames were manually labelled.

\noindent\textbf{Image pre-processing.} We used the same image preprocessing as in~\cite{Chaitanya_2019_IPMI}, where all images were bias corrected using N4 algorithm~\cite{Tustison_2010_N4ITK}. In addition, all images were centrally cropped into $128\times128$, given that the heart is generally located in the center of the image. This saves computational costs. 

% After that, for each volume data, intensities that fall in $2^{nd}$ and $98^{th}$ percentile were linearly re-scaled to $[0,1]$ and outliers were reset to its closest value (0 or 1). Then, all images were re-sampled to have a uniform intra-plane resolution of $1.367\times1.367mm^2$. Finally, 
% \noindent\textbf{Segmentation network.} For ease of comparison, same as \cite{Chaitanya_2019_IPMI}, we adopt the commonly-used 2D U-net as our segmentation network, which takes 2D image slices as input.
% Our framework supports various segmentation network architectures without any constraints. 

\noindent\textbf{Random data augmentation (Rand Aug).} We applied a strong random data augmentation method to our training data as a basic setting. Random affine transformation (i.e. scaling, rotation, translation), random horizontal and vertical flipping, random global intensity transformation (brightness and contrast)~\cite{Chaitanya_2019_IPMI} and elastic transformation were applied. 

\noindent\textbf{Training details.} For ease of comparison, same as \cite{Chaitanya_2019_IPMI}, we adopted the commonly-used 2D U-net as our segmentation network, which takes 2D image slices as input. The Adam optimizer with a batch size of 20 was used to update network parameters. For the proposed method, we first trained the network with the default data augmentation (Rand Aug) for 10,000 iterations (learning rate=$1e^{-3}$), and then finetuned the network by adding the proposed adversarial training using a smaller learning rate ($1e^{-5}$) for 2,000 iterations. The common standard cross-entropy loss function was used as $\SegLoss$. For bias field construction, we adopted the B-spline convolution kernel (order=3) with $4\times4$ control points. The kernel was provided by AirLab library~\cite{Sandkuhler_2018_Arxiv_Airlab}. We empirically set: $\alpha=0.3$, $\weight$ = 0.5, $\lambda_l=1$ and $\lambda_u=0.1$. Besides, we found that in our experiments, one step searching in the inner loop produced sufficient improvement. Thus, we set $n=1, \xi=1$ to save computational cost. All the experiments were performed on an Nvidia$^{\tiny{\text{\textregistered}}}$
GeForce$^{\tiny{\text{\textregistered}}}$ 2080 Ti with Pytorch. 
\subsection{Experiments and Results}
\label{SEC: results}
\noindent\textbf{Experiment 1: Low-shot learning.}
\label{sec: low-data experiment}
In this experiment, the proposed method was evaluated in both \emph{supervised} learning and \emph{semi-supervised} learning scenarios, where only 1 or 3 labelled subjects are available. Specifically, we used the same data splitting setting as in~\cite{Chaitanya_2019_IPMI}. The ACDC dataset was split into 4 subsets: a labelled set (where $N_l$ images were sampled from for training), unlabelled training set (N=25), validation set (N=2), test set (N=20). N denotes the number of subjects. Details of the low-data setting can be found in~\cite{Chaitanya_2019_IPMI}. For one-shot learning ($N_l$=1) and three-shot learning ($N_l$=3) in both supervised and semi-supervised settings, we trained the network for five times, each with a different labelled set.

We compared the proposed method (\textbf{Adv Bias}) with several competitive data augmentation methods including \textbf{VAT}~\cite{Miyato_2018_PAMI_VAT}, an effective data mixing-based method (\textbf{Mixup})~\cite{Zhang_2018_ICLR_mixup}  for supervised learning and the state-of-the-art semi-supervised generative model-based method(\textbf{cGANs})~\cite{Chaitanya_2019_IPMI}. For VAT and Mixup, we used the set of hyperparameters that achieved the best performance on the validation set and applied the same training procedure. For cGANs, we report the results of one-shot and three-shot learning in their original paper for reference, which were tested on the same test set. Table~\ref{tab:low_shot} compares the segmentation accuracy obtained by different data augmentation methods. Each reported value is the average Dice score of 20 test cases.
\begin{table}[t]
\begin{minipage}{.49\textwidth}
\caption{Comparison of the proposed method (Adv Bias) to other data augmentation methods.}
%Results with highest scores in each category are represented by bold font. }
\label{tab:low_shot}
\begin{adjustbox}{width=0.9\textwidth}
\begin{threeparttable}
\begin{tabular}{@{}lcrr@{}}
\toprule
\multirow{2}{*}{Setting} & \multicolumn{1}{c}{\multirow{2}{*}{Method}} & \multicolumn{2}{c}{\#  labelled subjects} \\
 & \multicolumn{1}{c}{} & 1 & 3 \\ \midrule
\multirow{5}{*}{Supervised} 
 & {No Aug} & 0.293 & 0.544 \\
& Rand Aug & 0.560 & 0.796 \\ \cmidrule(l){2-4} 
 & +Mixup\cite{Zhang_2018_ICLR_mixup} & 0.575 & 0.801 \\
 & +VAT\cite{Miyato_2018_PAMI_VAT} & 0.570 & 0.811 \\
 & +\textbf{Adv Bias } & \textbf{0.650} & \textbf{0.826} \\ \midrule
\multirow{3}{*}{Semi-supervised} & +VAT\cite{Miyato_2018_PAMI_VAT} & 0.625 & 0.826 \\
 & +\textbf{Adv Bias} & 0.692 & \textbf{0.830} \\ \cmidrule(l){2-4} 
 & cGANs\cite{Chaitanya_2019_IPMI} & \textbf{0.710} & 0.823 \\ \bottomrule
\end{tabular}
\end{threeparttable}
\end{adjustbox}
\end{minipage}
\hfill
\begin{minipage}{.48\textwidth}
\centering
\caption{Segmentation performance of the proposed method and baseline methods across five populations. All were trained with NOR cases only.}
\label{tab:cross_population}
\begin{adjustbox}{width=0.9\textwidth}
\begin{threeparttable}
\begin{tabular}{@{}lccccr@{}}
\toprule
Population & Rand Aug & +Mixup & +VAT & \begin{tabular}[c]{@{}c@{}}+\textbf{Adv Bias} \\ (Proposed)\end{tabular} \\  \midrule
{NOR} & 0.911 & 0.901 & 0.909 & \textbf{0.912} \\ \midrule
{ DCM} & 0.831 & 0.803 & 0.843 & \textbf{0.871} \\
{ HCM} & 0.871 & 0.881 & \textbf{0.891} & 0.890 \\
{ MINF} & 0.805 & 0.789 & 0.824 & \textbf{0.847} \\
{ ARV} & 0.843 & 0.844 & 0.843 &\textbf{ 0.853} \\ \midrule
Average & 0.841 & 0.833 & 0.853 &\textbf{ 0.868} \\ \bottomrule
\end{tabular}

% \begin{tablenotes}
% \item Reported values are average Dice scoresfor the left ventricular myocardium segmentation evaluated on five different populations. Results with highest scores are represented by bold font.
% \end{tablenotes}
\end{threeparttable}
\end{adjustbox}
\end{minipage}
\end{table}

In the supervised learning setting (no access to unlabelled images), when only one or three labelled subject was available, the proposed method clearly outperformed all baseline methods. For semi-supervised learning, the proposed methods outperformed VAT, especially when only one labelled subject is available (0.686 vs 0.625). The proposed method achieves competitive results compared to the semi-supervised GAN-based method (cGANs) as well. Of note, cGANs adopts two additional GANs to sample geometric transformations and intensity transformations from unlabelled images. This is why it was only compared in the semi-supervised learning setting here. On the contrary, our approach is applicable to both low-shot supervised learning and semi-supervised learning. In addition, cGANs contains more parameters than our method and thus it might be less computationally efficient.

\begin{figure}[t]
    \centering
    \includegraphics[width=0.5\textwidth]{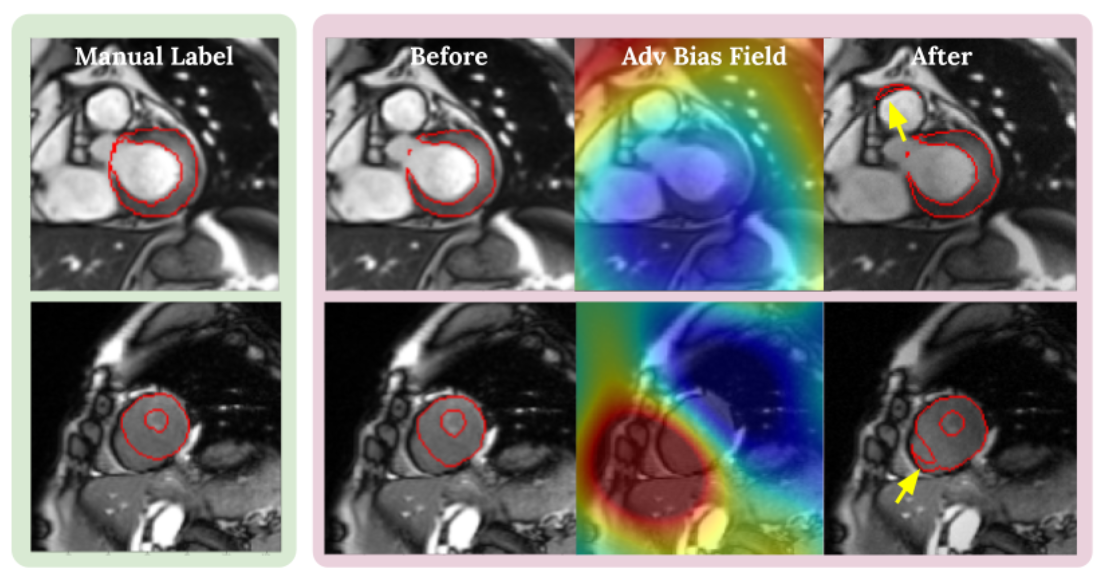}
    \caption{Visualization of generated adversarial examples and failed network predictions. Before/After: network prediction before/after bias field attack (Adv Bias Field).}
    \label{fig:adv_example}
\end{figure}

 %% Compared to GAN
% % However,  one limitation of generative model-based method their approach: 1) it requires two additional GANs with a large number of parameters, which are not computational efficient; 2) it is well known that training GANs can be hard and it is likely to have mode-collapse problem; 3) While the GAN-based method increases the geometric variations of training distributions, the anatomical shapes in these examples are not always realistic, as shown in the paper. One risk is that the output space distribution can be skewed by those abnormal shapes and the network may even produce unrealistic distorted shapes at test time.
\noindent\textbf{Experiment 2: Learning from limited population.}
\label{sec: limited population experiment}
In this experiment, we trained the network using only normal healthy subjects (NOR) and evaluated its performance on pathological cases. 20 healthy subjects were split into 14/2/4 subjects for training, validation and test. This setting simulates a practical data scarcity problem, where pathological cases are rarer, compared to healthy data. As shown in Table~\ref{tab:cross_population}, while the conventional method (Rand Aug) achieved excellent performance on the test healthy subjects (NOR), its performance dropped on pathological cases. Interestingly, applying Mixup did not help to solve this population shift problem, but rather slightly reduced the average performance compared to the baseline, from $0.841$ to $0.833$.  This might be due to the fact that Mixup generates unrealistic images through its linear combination of paired images, which may modify semantic features and affect representation learning for \emph{precise} segmentation. By contrast, our method outperformed both Mixup and VAT, yielding substantial and consistent improvements across five different populations. Notably, we attained evident improvement on the most challenging MINF images (0.805 vs 0.847), where the shape of the myocardium is clearly irregular. As shown in Fig.~\ref{fig:adv_example}, the proposed method does not only generate adversarial examples during training, but also increases the variety of image styles while preserving the shape information. Augmenting images with various styles can encourage the network to learn high-level shape-based representation instead of texture-based representation, leading to improved network robustness on unseen classes, as discussed in~\cite{Geirhos_2019_ICLR_BiasedTexure}. By contrast, VAT only introduces imperceptible noise, failing to model realistic image appearance variations. 

\noindent\textbf{Ablation study.} To get a better understanding of the effectiveness of adversarial bias field, we compared it to data augmentation using random bias field, using experiment setting 2. Results clearly showed that training with adversarial bias field improved the model generalization ability, increasing the Dice score from 0.852 to 0.868. On the other hand, applying $\CompositeDistance$ to regularize the network improved the average Dice score from 0.859 to 0.868, compared to the one trained with only $\KLDiv$.
Unlike random-based approach, constructing adversarial attacks considers both the posterior probability information estimated by the model and semantic information from images. In experiments, we found these attacks focused on attacking challenging images on which the network was uncertain, e.g. object boundary is not clear or there is another similar structure presented, see Fig.~\ref{fig:adv_example}. In the same spirit of online hard example mining, utilizing these borderline  examples during training helps the network to improve its generalization and robustness ability.
{Please find more details in the supplementary material}. 

\section{Discussion and Conclusion}
\label{SEC: conclusion}
In this work, we presented a realistic adversarial data augmentation method to improve the generalization and robustness for neural network-based medical image segmentation. We demonstrated that by modelling bias field and introducing adversarial learning, the proposed method is able to promote the learning of robust semantic features for cardiac image segmentation. It can also alleviate the data scarcity problem, as demonstrated in the low-data setting and cross-population experiments. The proposed method does not rely on generative networks but instead employs a small set of explainable and controllable parameters to augment data with image appearance variations which are realistic for MR. It can be easily extended for multi-class segmentation and used in general segmentation networks for improving model generalization and robustness.

% to also significantly increase the task complexity during training, as they can effectively deactivate semantic regions\todo{add figure to illustrate}.

% By regularizing the network produce consistent prediction regardless the changes, it is expected that the network can have better performance on unseen images.
% (2) The proposed method, as a data augmentation method, significantly increases image style variability of training data, which encourages the network to learn a high-level, shape-based representation rather than a low-level, texture-based representation~\cite{Geirhos_2019_ICLR_BiasedTexure}. 
\subsubsection{Acknowledgements.} 
This work was supported by the SmartHeart EPSRC Programme Grant (EP/P001009/1). HQ was supported by the EPSRC Programme Grant (EP/R005982/1).

\bibliographystyle{unsrt}
\bibliography{mybib}
\newpage
\end{document}

\section*{{Supplementary material}:~{Realistic Adversarial Data Augmentation for MR Image Segmentation}}

% https://www.miccai2020.org/en/PAPER-SUBMISSION-GUIDELINE.html#supplementary-material
% While the reviewers will have access to such supplementary material, they are under no obligation to review it, and the 8-page paper must contain all necessary information and illustrations by itself. Authors will be able to submit supplementary materials in the form of additional images, tables and proof of equations at the time of paper submission in the MICCAI submission format. These materials must not exceed two pages and must NOT bear any identification markers. Authors should not submit text materials beyond image and table captions, or definition of variables in equations. Captions should not exceed 100 words. Additionally, authors may submit supplementary videos without any identification markers. All supplementary material must be self-contained and zipped into a single file. Only the following formats are allowed: avi, doc, docx, mp4, pdf, wmv. We encourage authors to submit videos using an MP4 codec such as DivX contained in an AVI. Also, please submit a README text file with each video specifying the exact codec used and a URL where the codec can be downloaded.

% % If the paper is accepted, at the time of camera-ready version submission, authors may revise the supplementary material with format according to the guidelines above without the identification restrictions.

\setcounter{figure}{0}
\renewcommand{\figurename}{Supple. Figure}
\setcounter{table}{0}
\renewcommand{\tablename}{Supple. Table}

% [Would it be possible to generate the supplementary material using a separate latex file? To avoid you submitting the whole 12-page pdf onto CMT by accident.]

%\input{algorithms/adv_bias}
% \input{figures/adversarial_training}
\begin{figure}[!ht]
    \centering
    \includegraphics[width=\textwidth]{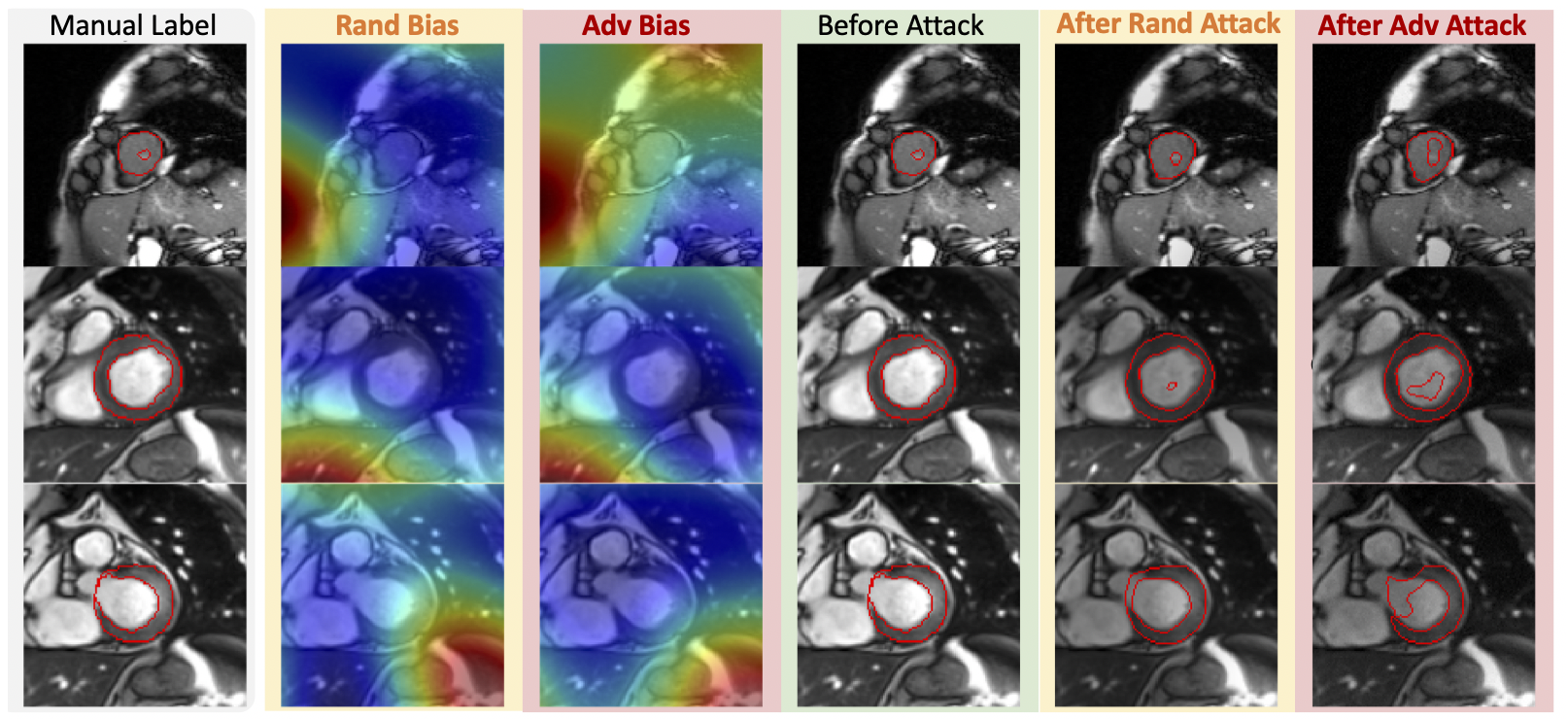}
    \caption{\textbf{Performance of  adversarial bias field attack
    (\textcolor{red}{ Adv Bias})
     vs random bias field attack (\textcolor{orange}{Rand Bias})}}
     %By comparing the predictions after rand/adv attacks (the last two columns) with the \textcolor{applegreen}{original predictions}, we can see that the network can be sensitive to the bias field perturbations. Interestingly, while the difference between the random and the adversarial bias field is mild, the proposed method is stronger at attacking the network. Therefore, adding these adversarial examples during training will encourage the network to learn more robust features for precise segmentation.} %Before/After: network prediction before/after bias field attack (Adv Bias).%}
    \label{fig:rand_vs_adv}
\end{figure}
\begin{figure}[!ht]
    \centering
    \includegraphics[width=0.5\textwidth]{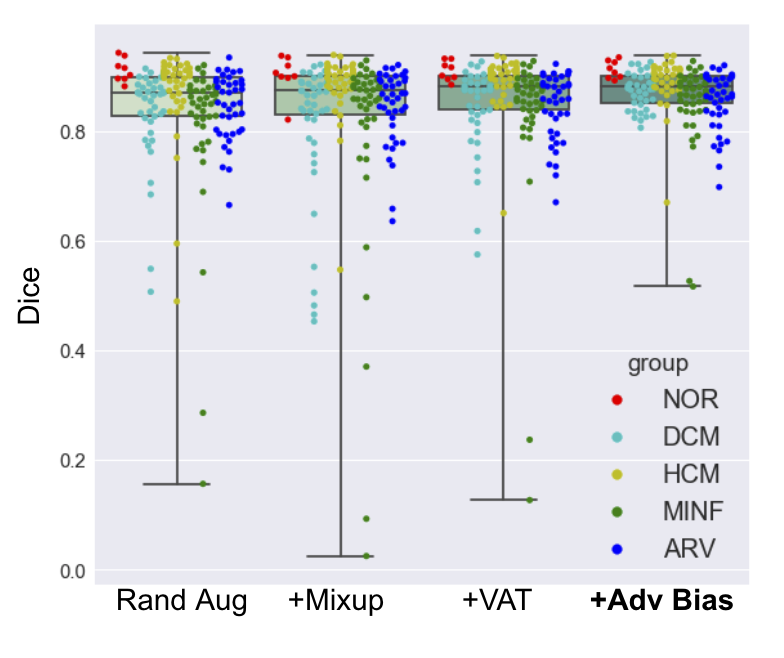}
    \caption{\textbf{Segmentation accuracy of each method for cardiac myocardium segmentation across five different populations.} Each dot represents the Dice score for each test subject, and its color indicates its group. Our method (column 4) produces more accurate segmentation on \textbf{unseen} pathological cases than the baselines. This indicates that the proposed method can improve the model robustness for abnormal cases, even the network was only trained with normal cases (NOR).}
    \label{fig:my_label}
\end{figure}

\begin{table}[!ht]
\centering
\caption{$\CompositeDistance$ vs $\KLDiv$}
\label{tab:Ablation study on distance function}
\begin{threeparttable}
\begin{tabular}{@{}ccccc@{}}
\toprule
Method & Distance Loss & Dice & HD & VolumeSim \\\midrule
VAT & $\KLDiv$ & 0.853 & 6.678 & 0.949 \\
VAT & $\CompositeDistance$ & 0.856 & 6.331 & 0.946 \\
Adv Bias & $\KLDiv$ & 0.859 & 6.330 & 0.949 \\
Adv Bias & $\CompositeDistance$ &\textbf{ 0.868} & \textbf{5.912} & \textbf{0.957} \\
\bottomrule
\end{tabular}
\begin{tablenotes}
\item  HD: Hausdorff distance; VolumeSim: Volume similarity index~\cite{Taha_2015_metric}. Reported values are average scores across all test subjects from five populations ($20\times4+4=84$ subjects). The same applies to Table~\ref{tab:Ablation study on adversarial optimization}.
\end{tablenotes}
\end{threeparttable}
\end{table}
\begin{table}[!htbp]
\centering
\caption{Random bias field vs Adversarial bias field}
\label{tab:Ablation study on adversarial optimization}
\begin{threeparttable}
\begin{tabular}{@{}ccccc@{}}
\toprule
Method & Distance Loss & Dice & HD & VolumeSim \\\midrule
Rand Bias&
$\CompositeDistance$ & 0.852 & 6.25 & 0.941 \\
Adv Bias & $\CompositeDistance$ & \textbf{0.868} & \textbf{5.91} & \textbf{0.957} \\
\bottomrule
\end{tabular}

\end{threeparttable}
\end{table}

% [The figure captions have more than 200 words, which seem to break the rule (https://www.miccai2020.org/en/PAPER-SUBMISSION-GUIDELINE.html). The reviewers may not have time to read carefully what each figure or table means. It would be sufficient to just have concise captions such as Performance of Proposed vs B or A vs B.]
\bibliographystyle{unsrt}
\bibliography{mybib}